\documentclass[prd,aps,floatfix,nofootinbib, 10 pt]{revtex4}
\usepackage{amssymb}
\usepackage{amsmath,graphicx,color,epsfig}

\begin{document}

\title{$\left( L_{e}-L_{\mu }-L_{\tau }\right) $ discrete symmetry for heavy
right-handed neutrinos and degenerate leptogenesis}
\author{Riazuddin}
\affiliation{Centre for Advanced Mathematics and Physics,\\
National University of Sciences and Technology,\\
Rawalpindi, Pakistan\\
and\\
National Centre for Physics,\\
Quaid-i-Azam University,\\
Islamabad, Pakistan.}

\begin{abstract}
The degenerate leptogenesis is studied when the degeneracy in two of the
heavy right-handed neutrinos [the third one is irrelevant if $\mu -\tau $
symmetry is assumed] is due to $\bar{L}\equiv \left( L_{e}-L_{\mu }-L_{\tau
}\right) $ discrete symmetry. It is shown that a sizeable leptogenesis
asymmetry $\left( \varepsilon \geq 10^{-6}\right) $ is possible. The
level of degeneracy required also predicts the Majorana phase needed for the
asymmetry and this prediction is testable since it is the same phase, which appears in the double $\beta $ decay and this prediction is testable. Implication of non-zero reactor
angle $\theta _{13}$ are discussed. It is shown that the contribution from $\sin ^{2}\theta _{13}$ to leptogenesis asymmetry parameter may even
dominate. An accurate measument of $\sin ^{2}\theta _{13}$ would have
important implications for the mass degeneracy of heavy right-handed
neutrinos.
\end{abstract}

\maketitle






\section{Introduction}

The purpose of this paper is to study degenerate(i.e. when two of the three
right handed neutrinos are (nearly) degenerate) leptogenesis in seesaw
mechanism where the mass matrix for right handed neutrinos has $\mu -\tau $
symmetry and the degeneracy is the result of $\bar{L}\equiv \left(
L_{e}-L_{\mu }-L_{\tau }\right) $ discrete symmetry. This is studied in a
generic seesaw gauge model \cite{1}, in which in addition to the usual
fermions and $SU_{L}(2)$ Higgs doublets, there are three $SU_{L}(2)$ singlet right handed neutrinos $N_{R}^{i}(i=e,\mu ,\tau )$ with $\mu -\tau $
symmetry and two Higgs with quantum numbers given below%
\begin{eqnarray}
L_{e} &:&(2,-1,0),\text{ }\phi ^{(1)}:(2,-1,0),\text{ }N_{R}^{e}\text{ :}%
(1,-1,1)  \notag \\
e_{R} &:&(1,-2,0)  \notag \\
L_{\mu -\tau } &:&(2,0,-1),\text{ }\phi ^{(2)}:(2,0,-1),\text{ }N_{R}^{\mu
,\tau }\text{ :}(1,1,-1)  \label{p1} \\
\mu _{R},\tau _{R} &:&(1,0,-2)  \notag \\
\Sigma &:&(1,0,0)  \notag \\
\Sigma ^{\prime } &:&(1,2,-2)  \notag
\end{eqnarray}
where the numbers in the parenthesis respectively correspond to $SU_{L}(2)$
and $U_{i}\left( 1\right) $ quantum numbers. It is important to remark that
as a result of $\mu -\tau $ symmetry the leptogenesis asymmetry parameter is
proportional to $\Delta m_{\text{sol}}^{2}$ \cite{2, 3} rather than $\Delta
m_{\text{atm}}^{2}$, and in general an unknown Majorana phase which,
however, also appears in the neutrinoless double $\beta -$decay. This was
studied for the hierarchal $\left( M_{2}\gg M_{1}\right) $ leptogenesis. Now
a study is made for degenerate leptogenesis when the degeneracy is the
result of $\bar{L}$ discrete symmetry for the right-handed heavy neutrinos
sector.This degeneracy is protected by the symmetry (although global) and as
such would \ be softly broken. It is shown that a sizeable lepton asymmetry
\ $\left( \varepsilon \geq 10^{-6}\right) $ is possible. The level of
degeneracy needed for this to occur also predicts the Majorana phase needed
for the asymmetry. This is the distinguishing feature of the model
considered. In general the asymmetry parameter is proportional to the
product of degeneracy parameter $\left( \frac{\Delta M}{M}\right) ^{-1}$ and
a CP-violating phase; the fixation of this product to get a sizable
leptogenesis parameter does not necessarily predict one from the other. This
is because they do not get related, in contrast to the model considered
here, see for example \cite{4, 5}. Since the phase involved is the same
which occurs in the neutrinoless double $\beta -$decay, this prediction is
testable. Further the effect of a non-zero reactor angle $\theta _{13}$ on
leptogenesis is considered in some detail. It is shown that the contribution
from $\sin ^{2}\theta _{13}$ to leptogensis asymmetry parameter may even
dominate. As such an accurate measurement of $\sin ^{2}\theta _{13}$ would
have important implications for the mass spectrum of heavy right-handed
neutrinos, particularly for $\frac{M_{2}-M_{1}}{M_{2}+M_{1}}$.

The Yukawa couplings of neutrinos with Higgs, using $\mu -\tau $ symmetry
for right-handed neutrinos only, is given by
\begin{eqnarray}
\mathcal{L}_{Y} &=&g_{11}\bar{L}_{e}e_{R}\tilde{\phi}^{(1)}+\left[ g_{22}%
\bar{L}_{\mu }\mu _{R}+g_{23}\bar{L}_{\mu }\tau _{R}+g_{32}\bar{L}_{\tau
}\mu _{R}+g_{33}\bar{L}_{\tau }\tau _{R}\right] \tilde{\phi}^{(2)}+hc  \notag
\\
&&+h_{11}\overline{L}_{e}N_{e}\phi ^{(2)}+[h_{22}\overline{L}_{\mu }(N_{\mu
}+N_{\tau })+h_{32}\overline{L}_{\tau }(N_{\mu }+N_{\tau })]\phi ^{(1)}
\label{p2} \\
&&+hc+f_{11}N_{e}^{T}CN_{e}\Sigma ^{\prime }+f_{12}N_{e}^{T}C(N_{\mu
}+N_{\tau })\Sigma +hc  \notag \\
&&+[f_{22}(N_{\mu }^{T}CN_{\mu }+N_{\tau }^{T}CN_{\tau })+f_{23}(N_{\mu
}^{T}CN_{\tau }+N_{\tau }^{T}CN_{\mu })]\overline{\Sigma }^{\prime }  \notag
\end{eqnarray}%
where
\begin{equation*}
\tilde{\phi}=-i\tau _{2}\phi ^{\ast },\,\phi =\left(
\begin{array}{l}
\phi ^{0} \\
-\phi ^{-}%
\end{array}%
\right) ,\,\tilde{\phi}=\left(
\begin{array}{l}
\phi ^{+} \\
\phi ^{0}%
\end{array}%
\right)
\end{equation*}
One can also write Yukawa couplings of quarks with Higgs doublets, in the
same fashion as for charged lepton as follows:%
\begin{eqnarray}
L_{Y} &=&G_{11}\bar{L}_{u}u_{R}\phi ^{(1)}+G_{22}\bar{L}_{c}c_{R}\varphi
^{(2)}+G_{33}\bar{L}_{t}t_{R}\phi ^{(2)}+\tilde{G}_{11}\bar{L}_{u}d_{R}%
\tilde{\phi}^{(1)}  \label{modify-1} \\
&&+\left( \tilde{G}_{22}\bar{L}_{c}s_{R}+\tilde{G}_{23}\bar{L}_{c}b_{R}+%
\tilde{G}_{32}\bar{L}_{t}s_{R}+\tilde{G}_{33}\bar{L}_{t}b_{R}\right) \tilde{%
\phi}^{(2)}  \notag
\end{eqnarray}
A remark about Yukawa Couplings is in order. The model contains two Higgs
doublets $\phi ^{(1)}$ and $\phi ^{(2)}$, the former is coupled to the first
generation while the later to second and third generations. Then the quantum
number given in Eq. (\ref{p1}) dictate the couplings as in Eq. (\ref{p2})
and (\ref{modify-1}). Except for heavy $SU\left( 2\right) $ singlet right
handed neutrinos, $2\leftrightarrow 3$ symmetry is not the symmetry of the
Lagrangian in Eq. (\ref{p2}). The Yukawa couplings with quarks will not be
considered further as they are not relevant for what follows. In general the
Yukawa couplings used above are complex. It is convenient to introduce
Yukawa coupling matrices $Y_{l}$ and $Y_{D}$.
\begin{equation}
Y_{l}=\left(
\begin{array}{ccc}
g_{11} & 0 & 0 \\
0 & g_{22} & g_{23} \\
0 & g_{32} & g_{33}%
\end{array}%
\right) \text{, }Y_{D}=\left(
\begin{array}{ccc}
h_{11} & 0 & 0 \\
0 & h_{22} & h_{23} \\
0 & h_{32} & h_{33}%
\end{array}%
\right)   \label{insertion3}
\end{equation}%
Then the charged lepton and Dirac neutrino mass matrices are
\begin{equation}
M_{l}=\left(
\begin{array}{lll}
g_{11}v_{1} & 0 & 0 \\
0 & g_{22}v_{2} & g_{23}v_{2} \\
0 & g_{32}v_{2} & g_{33}v_{2}%
\end{array}%
\right)   \label{insertion5}
\end{equation}%
\begin{equation}
m_{D}=\left(
\begin{array}{lll}
h_{11}v_{2} & 0 & 0 \\
0 & h_{22}v_{1} & h_{22}v_{1} \\
0 & h_{32}v_{1} & h_{32}v_{1}%
\end{array}%
\right)   \label{insertion6}
\end{equation}%
while $M_{R}$ is
\begin{equation}
M_{R}=\left(
\begin{array}{lll}
f_{11}\Lambda ^{\prime } & f_{12}\Lambda  & f_{12}\Lambda  \\
f_{12}\Lambda  & f_{22}\Lambda ^{\prime } & f_{23}\Lambda ^{\prime } \\
f_{12}\Lambda  & f_{23}\Lambda ^{\prime } & f_{22}\Lambda ^{\prime }%
\end{array}%
\right)   \label{insertion7}
\end{equation}%
where $\left\langle \phi _{1,2}\right\rangle =v_{1,2}$, $\left\langle \Sigma
\right\rangle =\Lambda $, $\left\langle \Sigma ^{\prime }\right\rangle
=\Lambda ^{\prime }$. It is convenient to have a basis in which $Y_{l}$ and $%
M_{R}$ are simultaneously diagonal:
\begin{equation}
Y_{l}\rightarrow \hat{Y}_{l}=U_{L}^{-1}Y_{l}U_{E}.  \label{insertion8}
\end{equation}%
Correspondingly [for left-handed doublets $L_{i}$ and right handed singlets $%
E_{iR}$]
\begin{equation}
L_{i}\rightarrow U_{L}L_{i}\text{, }E_{iR}\rightarrow U_{R}E_{iR}
\label{insertion9}
\end{equation}%
where $i=e,\,\mu ,\,\tau $ is the flavor index.

It is pertinent to remark that by imposing the $\mu -\tau $ symmetry at the
Lagrangian level only on the $SU\left( 2\right) $ singlet right handed
neutrinos, a well known problem \cite{6, 7} for simultaneous imposing of $%
\mu -\tau $ symmetry on the left handed charged leptons and the left handed
neutrinos, is avoided. For example, in the basis where charged leptons are
diagonal, this would imply $m_{\mu }=m_{\tau }$. $\nu _{\mu }-\nu _{\tau }$
symmetry can, however be imposed on $m_{D}$ independent of the $\mu -\tau $
symmetry for the right handed neutrinos, giving $h_{22}=h_{33}$ implying in
turn the maximum atmosphere mixing angle and $\theta _{13}=0$. We would not
impose this symmetry exactly and in fact consider consequence of its
breaking, in particular $\theta _{13}\neq 0$, which have important
implications for leptogenesis asymmetry parameter. It, may in fact, dominate
depending on the value of $\sin ^{2}\theta _{13}.$ Thus an accurate
measurement $\sin ^{2}\theta _{13}$ would be of great interest.

\section{Mass Matrices in Seesaw Mechanism with $\bar{L}$ Discrete
symmetry}

As is well known $M_{R}$ as given in Eq. (\ref{insertion7}) is diagonalized
by a mixing matrix with $\sin ^{2}\theta _{23}^{\prime }=\frac{1}{2}$ and $%
\theta _{13}^{\prime }=0$, i.e. by
\begin{equation}
V=\left(
\begin{array}{lll}
\cos \theta _{12}^{\prime } & \sin \theta _{12}^{\prime } & 0 \\
-\frac{\sin \theta _{12}^{\prime }}{\sqrt{2}} & \frac{\cos \theta
_{12}^{\prime }}{\sqrt{2}} & -\frac{1}{\sqrt{2}} \\
-\frac{\sin \theta _{12}^{\prime }}{\sqrt{2}} & \frac{\cos \theta
_{12}^{\prime }}{\sqrt{2}} & \frac{1}{\sqrt{2}}%
\end{array}%
\right) P(\gamma )  \label{insertion10}
\end{equation}%
where $P(\gamma )$ is diagonal phase matrix (consisting of non-trivial
Majorana phases $\gamma _{1}$, $\gamma _{2}$, $\gamma _{3}$). Thus
\begin{equation}
V^{T}M_{R}V=\hat{M}_{R}=diag\left( \hat{M}_{1},\hat{M}_{2},\hat{M}_{3}\right)
\label{insertion11}
\end{equation}%
with $\hat{M}_{i}=M_{i}e^{2i\gamma _{i}}$, $i=1,2,3$ and
\begin{equation}
\tan 2\theta _{12}^{\prime }=\frac{2\sqrt{2}f_{12}\Lambda }{%
(f_{22}+f_{23}-f_{11})\Lambda ^{\prime }}  \label{insertion12}
\end{equation}%
\begin{eqnarray}
\hat{M}_{3} &=&M_{3}e^{2i\gamma _{3}}  \label{insertion13} \\
&=&\left[ f_{22}-f_{23}\right] \Lambda ^{\prime }e^{2i\gamma _{3}}  \notag
\end{eqnarray}%
Then the effective Majorana mass matrix for the light neutrinos is
\begin{equation}
M_{\nu }=\hat{m}_{D}\hat{M}_{R}^{-1}\hat{m}_{D}^{T}  \label{insertion14}
\end{equation}%
where $\hat{m}_{D}$ is the Dirac matrix in $\left(
\begin{array}{lll}
\bar{N}_{1} & \bar{N}_{2} & \bar{N}_{3}%
\end{array}%
\right) \left(
\begin{array}{l}
\nu _{e} \\
\nu _{\mu } \\
\nu _{\tau }%
\end{array}%
\right) $basis:
\begin{equation*}
\hat{m}_{D}=m_{D}V^{\ast }
\end{equation*}%
and the corresponding Yukawa matrix is
\begin{equation}
\hat{Y}_{D}=Y_{D}V^{\ast }  \label{insertion15}
\end{equation}

Before proceeding further, let me display the Higgs potential for $\phi $
fields
\begin{eqnarray}
V_{H} &=&\mu _{1}^{2}\left\vert \phi _{1}\right\vert ^{2}+\mu _{2}\left\vert
\phi _{2}\right\vert ^{2}+\mu _{3}\left( \phi _{1}^{\dagger }\phi _{2}+\phi
_{2}^{\dagger }\phi _{1}\right) +\lambda _{1}\left( \phi _{1}^{\dagger }\phi
_{1}\right) ^{2}  \notag \\
&&+\lambda _{2}\left( \phi _{2}^{\dagger }\phi _{2}\right) ^{2}+\lambda
_{3}\left( \phi _{1}^{\dagger }\phi _{2}+\phi _{2}^{\dagger }\phi
_{1}\right) ^{2}-\lambda _{4}\left( \phi _{1}^{\dagger }\phi _{2}-\phi
_{2}^{\dagger }\phi _{1}\right) ^{2}  \label{insertion16} \\
&&+\lambda _{5}\left( \phi _{1}^{\dagger }\phi _{1}\right) \left( \phi
_{2}^{\dagger }\phi _{2}\right)  \notag
\end{eqnarray}%
When the symmetry is broken
\begin{equation}
\tilde{\phi}_{i}=\left(
\begin{array}{l}
H_{i}^{+} \\
v_{i}+h_{i}+ia_{i}
\end{array}%
\right) ,\,i=1,2.  \label{insertion17}
\end{equation}%
Due to the presence of the terms $\lambda _{3}$ and $\lambda _{4}$ in Eq. (%
\ref{insertion16}), when the symmetry is broken, each pair of Higgs
particles $\left( h_{1},h_{2}\right) ,\,\left( a_{1},a_{2}\right) $ and $%
\left( H_{1}^{+},H_{2}^{+}\right) $ mix. Further, when the resulting mass
matrices are diagonalized, one of the charged Higgs and one of the neutral
Higgs acquire zero masses and as such are eaten up by $W^{+}$ and $Z^{0}$ to
give them masses. As a result one has four massive physical Higgs particles,
one charged $H^{+}$ and three neutral $H$, $h$ and $A^{0}$. Due to presence
of Majorana mass term
\begin{equation}
H_{M}=N^{T}CM_{R}N+h.c.  \label{insertion18}
\end{equation}%
providing explicit breaking of family lepton number, flavor changing
interactions can arise due to radiative corrections and are controlled by
elements of $M_{R}$ and have been shown to be calculable and finite\cite{8}.
At one loop level such corrections arise due to charged Higgs and have been
shown \cite{8} to be highly suppressed. It may be remarked here that the
Higgs potential for $\Sigma $ fields can be included but even after breaking
of the symmetry there is no mixing between $\Sigma $ and $\phi$ fields. $%
\Sigma ^{\prime }$ gives mass to one of the neutral gauge bosons and make it
super heavy. $\Sigma $ is not coupled to gauge bosons. But both $\Sigma $
and $\Sigma ^{\prime }$ give mass to right handed neutrinos.

We now apply $\bar{L}$ discrete symmetry on the purely heavy right-handed
neutrino part of the Lagrangian (\ref{p2}) i.e.%
\begin{equation}
N_{i}\rightarrow e^{i\xi L}N_{i}  \label{modify2}
\end{equation}%
which leaves only the $f_{12}$ term invariant so that%
\begin{equation}
f_{11}=0=f_{22}=f_{33}  \label{insertion19}
\end{equation}%
As a result%
\begin{equation}
\hat{M}_{1}=Me^{2i\gamma _{1}}\text{, }\hat{M}_{2}=-Me^{2i\gamma
_{2}}=Me^{2i\gamma _{2}^{\prime }}  \label{inseration20}
\end{equation}%
where $M=\sqrt{2}\left\vert f_{12}\Lambda \right\vert $, $\gamma
_{2}^{\prime }=\gamma _{2}+\frac{\pi }{2}$ so that the minus sign in Eq. (%
\ref{inseration20}) has been absorbed in the redefinition of the Majrona
phase $\gamma _{2}$. Further $\theta _{12}^{\prime }=\pm \frac{\pi }{4}$. To
break the $\bar{L}$ discrete symmetry so as to obtain nearly degeneracy of $M_{1}$ and $M_{2}$, we assume $\left\vert f_{22}\Lambda ^{\prime
}\right\vert $, $\left\vert f_{23}\Lambda ^{\prime }\right\vert $, $\left\vert f_{11}\Lambda ^{\prime }\right\vert \leq \left\vert f_{12}\Lambda
\right\vert $ and $f_{22}\simeq f_{23}$ so that the third right-handed
neutrino becomes sterile with $M_{3}\simeq $ a few eV. Then it is easy to see
that $\left[ \Delta M=\frac{M_{2}-M_{1}}{2}\text{, }M=\frac{M_{2}+M_{1}}{2}
\right] $
\begin{equation}
\frac{\Delta M}{M}=\left\vert \eta \right\vert \text{, }\tan \theta
_{12}^{\prime }=\pm 1-\eta ^{\prime }  \label{insertion21}
\end{equation}
where%
\begin{equation}
\eta =\frac{\left( f_{22}+f_{23}+f_{11}\right) \Lambda ^{\prime }}{2\sqrt{2}%
f_{12}\Lambda }\text{, }\eta ^{\prime }=\frac{\left(
f_{22}+f_{23}-f_{11}\right) \Lambda ^{\prime }}{2\sqrt{2}f_{12}\Lambda }
\label{insertion22}
\end{equation}
The degree of degeneracy needed for providing sizable asymmetry [see section
3] requires $\left\vert \eta \right\vert \simeq 10^{-3}$ and correspondingly
$\eta ^{\prime }$ is also of the same order. A remark about the sterile
neutrino would be in order. Even if $\bar{L}$ discrete symmetry is broken,
the sterile neutrino can not mix with any of the active neutrinos unless $%
\mu -\tau $ symmetry is also broken for right handed neutrinos \cite{1}.
Even then the primordial nucleosynthesis bound on the active member of
neutrino at $t\sim 1s$: $N_{\nu }< 3.1$ implies that the
oscillation of active neutrinos into the sterile one should obey the bound $%
\delta m^{2}\sin ^{2}2\theta \leq 1.6\times 10^{-6}$eV$^{2}$ which excludes
the $\nu _{\mu }\rightarrow \nu _{s}$ and $\nu _{e}\rightarrow \nu _{s}$
oscillations and as such do not effect the atmospheric and solar neutrino
solutions \cite{9}.

The effective Majorana mass matrix for light neutrinos, given in Eq. (\ref{insertion14}), is
\begin{equation}
M_{\nu }=\widehat{m}_{D}\widehat{M}_{R}^{-1}\widehat{m}_{D}^{T}=\widehat{A}
\label{p19}
\end{equation}
where $\widehat{A}$ is $3\times 3$ matrix with matrix elements%
\begin{eqnarray}
a_{11} &=&h_{11}^{2}v_{2}^{2}A  \notag \\
\sqrt{2}a_{12} &=&h_{11}(2h_{22})v_{1}v_{2}B  \notag \\
\sqrt{2}a_{13} &=&h_{11}(2h_{32})v_{1}v_{2}B  \label{p20} \\
a_{22} &=&\frac{1}{2}(4h_{22}^{2}v_{1}^{2})C  \notag \\
a_{23} &=&\frac{1}{2}(2h_{22})(2h_{32})v_{1}^{2}C  \notag \\
a_{33} &=&\frac{1}{2}(4h_{32}^{2})v_{1}^{2}C  \notag
\end{eqnarray}
Here
\begin{eqnarray}
A &=&\frac{e^{-i(\gamma _{1}+\gamma _{2}^{\prime })}}{M}\left\{ \cos \frac{%
\Delta \gamma }{2}-i\frac{\Delta M}{M}\sin \frac{\Delta \gamma }{2}\right\}
=C  \notag \\
B &=&-\frac{e^{-i(\gamma _{1}+\gamma _{2}^{\prime })}}{M}c^{\prime
}s^{\prime }\left\{ \frac{\Delta M}{M}\cos \frac{\Delta \gamma }{2}-i\sin
\frac{\Delta \gamma }{2}\right\}  \label{p21-a}
\end{eqnarray}%
where%
\begin{eqnarray}
\Delta \gamma &=&2(\gamma _{1}-\gamma _{2}^{\prime })  \notag \\
c^{\prime }s^{\prime } &=&\pm \frac{1}{2}  \label{p22a}
\end{eqnarray}%
If one assumes $\nu _{\mu }\rightarrow \nu _{\tau }$ symmetry for the $%
M_{\nu }$, then $h_{22}=h_{32}$ would imply, as is well known, maximal $%
\theta _{23}=\pm \pi /4$, $\theta _{13}=0$ and $m_{3}=0$. If $\theta _{23}$
is not exactly maximal, or $\theta _{13}\neq 0$, then $\nu _{\mu
}\rightarrow \nu _{\tau }$ symmetry for left-handed neutrinos is broken but $%
m_{3}$ is still zero since the second and third columns of $m_{D}$ \ given
in Eq. (\ref{insertion6}) are identical \cite{6}. However, present
experiments indicate that the breaking has to be small. Thus defining $%
h_{\pm }=\frac{h_{22}\pm h_{33}}{2}$, where $\left\vert \frac{h_{+}}{h_{-}}%
\right\vert \ll 1$, we have, neglecting $\left\vert \frac{h_{+}}{h_{-}}%
\right\vert ^{2}$,%
\begin{eqnarray}
\sqrt{2}\left( a_{12}-a_{13}\right) &=&\left( h_{11}v_{2}\right) \left(
2h_{+}v_{1}\right) \left[ 2\frac{h_{-}}{h_{+}}\right] B  \notag \\
\left( a_{22}-a_{23}\right) &=&\frac{1}{2}\left( 4h_{+}^{2}v_{1}^{2}\right) %
\left[ 4\frac{h_{-}}{h_{+}}\right] C  \notag \\
a_{23} &=&\frac{1}{2}\left( 4h_{+}^{2}v_{1}^{2}\right) C  \label{modify3}
\end{eqnarray}%
To quantify the breaking and in order not to introduce too many parameters,
we assume the maximal atmospheric angle, but $\theta _{13}\neq 0$. The the $%
M_{\nu }$ as given in Eq. (\ref{p19}) can be diagonalized with the matrix%
\cite{6}%
\begin{equation}
U=\left(
\begin{array}{ccc}
c & s & s_{2} \\
-\frac{s-s_{2}}{\sqrt{2}} & \frac{c+s_{2}}{\sqrt{2}} & -\frac{1}{\sqrt{2}}
\\
-\frac{s+s_{2}}{\sqrt{2}} & \frac{c-s_{2}}{\sqrt{2}} & \frac{1}{\sqrt{2}}%
\end{array}%
\right) \times diag\left( e^{i\beta _{1}},e^{i\beta _{2}},e^{i\beta
_{3}}\right)  \label{modify4}
\end{equation}%
where $c=\cos \theta _{12}$, $s=\sin \theta _{12}$, $\theta _{12}$ is solar
mixing angle and $s_{2}=\sin \theta _{13}$, $\theta _{13}$ is the reactor
angle. Finally, then the elements of $M_{\nu }$ are%
\begin{eqnarray}
a_{11} &=&e^{-i\left( \beta _{1}+\beta _{2}\right) }m\left\{ \cos \frac{%
\Delta }{2}\left( 1-\frac{\Delta m}{m}\cos 2\theta _{12}\right) -i\sin \frac{%
\Delta }{2}\left( \cos 2\theta _{12}-\frac{\Delta m}{m}\right) \right\}
\notag \\
\sqrt{2}a_{12\left( 13\right) } &=&e^{-i\left( \beta _{1}+\beta _{2}\right)
}m\left\{ \sin 2\theta _{12}\left[ \frac{\Delta m}{m}\cos \frac{\Delta }{2}%
+i\sin \frac{\Delta }{2}\right] \pm s_{2}\left( \left( 1-\cos 2\theta _{12}%
\frac{\Delta m}{m}\right) \cos \frac{\Delta }{2}+i\left( \frac{\Delta m}{m}%
-\cos 2\theta _{12}\right) \sin \frac{\Delta }{2}\right) \right\}  \notag \\
2a_{22\left( 33\right) } &=&e^{-i\left( \beta _{1}+\beta _{2}\right)
}m\left\{ \cos \frac{\Delta }{2}\left( 1+\frac{\Delta m}{m}\cos 2\theta
_{12}\right) +i\sin \frac{\Delta }{2}\left( \frac{\Delta m}{m}+\cos 2\theta
_{12}\right) \pm 2s_{2}\sin 2\theta _{12}\left( \frac{\Delta m}{m}\cos \frac{%
\Delta }{2}+i\sin \frac{\Delta }{2}\right) \right\}  \label{insertion26} \\
a_{23} &=&e^{-i\left( \beta _{1}+\beta _{2}\right) }m\left\{ \cos \frac{%
\Delta }{2}\left( 1+\frac{\Delta m}{m}\cos 2\theta _{12}\right) +i\sin \frac{%
\Delta }{2}\left( \frac{\Delta m}{m}+\cos 2\theta _{12}\right) \right\}
\notag \\
\Delta &=&2(\beta _{1}-\beta _{2})  \notag \\
~m &=&\frac{m_{2}+m_{1}}{2},~\Delta m=\frac{m_{2}-m_{1}}{2}.  \notag
\end{eqnarray}%
Some combinations of the above parameters are needed to calculate the
asymmetry parameter in leptogenesis, which are now summarized. Calculation
of $\Im\left[ 2a_{12}a_{13}a_{11}^{\ast }2a_{23}^{\ast }\right] $ from Eqs. (%
\ref{p20}) and (\ref{insertion26}) and equating them, gives%
\begin{eqnarray}
c^{\prime 2}s^{\prime 2}\sin [2(\gamma _{1}-\gamma _{2}^{\prime })]\frac{%
M_{2}^{2}-M_{1}^{2}}{M_{1}^{3}M_{2}^{3}} &=&-\frac{1}{\left\vert
h_{11}v_{2}\right\vert ^{4}\left\vert 2h_{22}v_{1}\right\vert ^{2}\left\vert
2h_{32}v_{2}\right\vert ^{2}}m_{1}m_{2}\times  \label{27b} \\
&&\left[ c^{2}s^{2}(m_{2}^{2}-m_{1}^{2})+\frac{1}{2}s_{2}^{2}\cos 2\theta
_{12}\left(
\begin{array}{c}
m_{1}^{2}+m_{2}^{2} \\
-8m_{1}m_{2}c^{2}s^{2}\sin ^{2}\frac{\Delta }{2}-2c^{2}s^{2}\left(
m_{1}-m_{2}\right) ^{2}%
\end{array}%
\right) \right] \sin \Delta .  \notag
\end{eqnarray}%
Further calculating $\left\vert 2a_{11}a_{23}-2a_{12}a_{13}\right\vert $
from the same equations and equating them gives%
\begin{equation}
m_{1}m_{2}\left[ 1+O\left( s_{2}^{2}\right) \right] =\frac{1}{M_{1}M_{2}}%
\left[ \left\vert h_{11}v_{2}\right\vert ^{2}\left\vert
2h_{22}v_{1}\right\vert \left\vert h_{32}v_{2}\right\vert \right]
\label{modify5}
\end{equation}%
Another useful relation is obtained by calculating $\left\vert
2a_{12}a_{13}\right\vert $ from equations (\ref{p20}) and (\ref{insertion26}%
) and equating them%
\begin{eqnarray}
&&\left\vert c^{2}s^{2}\left( (m_{2}^{2}+m_{1}^{2})\cos \Delta
-2m_{1}m_{2}\right) +i\sin \Delta \left(
c^{2}s^{2}(m_{2}^{2}-m_{1}^{2})+s_{2}^{2}\cos 2\theta
_{12}(m_{2}^{2}+m_{1}^{2})\right) \right\vert  \notag \\
&=&c^{\prime 2}s^{\prime 2}\frac{m_{1}m_{2}}{M_{1}M_{2}}\left[ \left(
M_{2}-M_{1}\right) ^{2}+4M_{1}M_{2}\sin ^{2}\left( \gamma _{1}-\gamma
_{2}^{\prime }\right) \right]  \label{modify6}
\end{eqnarray}%
where terms of order $s_{2}^{2}$ compared to $1$ and $\frac{c^{2}s^{2}}{%
1-2c^{2}s^{2}}$ and of order $s_{2}^{2}\left( m_{2}^{2}-m_{1}^{2}\right) $
compared to $\left( m_{2}^{2}+m_{1}^{2}\right) $ have been neglected and Eq.
(\ref{modify5}) has been used. This gives, on neglecting terms of order $%
s_{2}^{4}\left( \frac{\Delta m}{m}\right) ^{2}$, $s_{2}^{8}$ and $%
s_{2}^{6}\left( \frac{\Delta m}{m}\right) ,$%
\begin{eqnarray}
&&\left\{ c^{2}s^{2}\left[ (m_{2}-m_{1})^{2}+4m_{1}m_{2}\sin ^{2}\frac{%
\Delta }{2}\right] +\frac{\cos 2\theta _{12}}{c^{2}s^{2}}\frac{%
m_{2}^{2}+m_{1}^{2}}{2m_{1}m_{2}}s_{2}^{2}\left[
2c^{2}s^{2}(m_{2}^{2}-m_{1}^{2})+s_{2}^{2}\cos 2\theta _{12}\left(
m_{2}^{2}+m_{1}^{2}\right) \right] \right\}  \notag \\
&=&c^{\prime 2}s^{\prime 2}\frac{m_{1}m_{2}}{M_{1}M_{2}}\left[ \left(
M_{2}-M_{1}\right) ^{2}+4M_{1}M_{2}\sin ^{2}\left( \gamma _{1}-\gamma
_{2}^{\prime }\right) \right]  \label{modify7}
\end{eqnarray}%
In the present case, when $c^{\prime 2}s^{\prime 2}=\frac{1}{4}$, the
relations (\ref{27b}) [on using Eq.(\ref{modify5}) and (\ref{modify7})]
become [$\Delta _{\gamma }=2\left( \gamma _{1}-\gamma _{2}^{\prime }\right) $]
\begin{equation}
\sin \Delta _{\gamma }=-\sin ^{2}2\theta _{12}\frac{\Delta m/m}{\Delta M/M}%
\sin \Delta \left[ \left( 1+\frac{1}{\sin ^{2}2\theta _{12}}\left( 1-\sin
^{2}2\theta _{12}\sin ^{2}\frac{\Delta }{2}\right) \right) \cos 2\theta
_{12}r^{\prime }\right]  \label{modify8}
\end{equation}%
\begin{equation}
\sin ^{2}2\theta _{12}\left[ \sin ^{2}\frac{\Delta }{2}+\left( \Delta
m/m\right) ^{2}\left( 1+8\frac{\cos 2\theta _{12}}{\sin ^{4}2\theta _{12}}%
r^{\prime }\left( \sin ^{2}2\theta _{12}+r^{\prime }\cos 2\theta
_{12}\right) \right) \right] =\left[ \left( \Delta M/M\right) ^{2}+\sin ^{2}%
\frac{\Delta _{\gamma }}{2}\right]  \label{modify9}
\end{equation}%
where%
\begin{equation*}
r^{\prime }=s_{2}^{2}/\left( \Delta m/m\right) =\sin ^{2}\theta _{13}/\left(
\Delta m/m\right)
\end{equation*}%
Now the Yukawa couplings $\left\vert h_{11}v_{2}\right\vert $, $\left\vert
2h_{+}v_{1}\right\vert $ and $\left\vert \frac{h_{-}}{h_{+}}\right\vert $
can be evaluated. From Eqs. (\ref{p20}), (\ref{modify3})%
\begin{eqnarray}
\left\vert h_{11}v_{2}\right\vert ^{2} &=&\frac{\left\vert a_{11}\right\vert
}{\left\vert C\right\vert }\text{, }\left\vert 2h_{+}v_{1}\right\vert ^{2}=2%
\frac{\left\vert a_{23}\right\vert }{\left\vert C\right\vert }  \notag \\
\left\vert 2h_{+}\right\vert \left\vert 2h_{-}\right\vert v_{1}^{2} &=&\frac{%
\left\vert a_{22}-a_{33}\right\vert }{2\left\vert C\right\vert }
\label{27c2}
\end{eqnarray}
From Eqs. (\ref{insertion26})
\begin{eqnarray}
\left\vert a_{11}\right\vert &=&m\left[ 1-\sin ^{2}2\theta _{12}\sin
^{2}\left( \frac{\Delta }{2}\right) -2\cos 2\theta _{12}\frac{\Delta m}{m}%
+O\left( \left( \frac{\Delta m}{m}\right) ^{2}\right) \right] ^{1/2}  \notag
\\
\left\vert a_{23}\right\vert &=&\frac{m}{2}\left[ 1-\sin ^{2}2\theta
_{12}\sin ^{2}\left( \frac{\Delta }{2}\right) +2\cos 2\theta _{12}\frac{%
\Delta m}{m}+O\left( \left( \frac{\Delta m}{m}\right) ^{2}\right) \right]
^{1/2}  \notag \\
\left\vert a_{22}-a_{33}\right\vert &=&2ms_{2}\sin 2\theta _{12}\left[ \sin
^{2}\left( \frac{\Delta }{2}\right) +\left( \frac{\Delta m}{m}\right) ^{2}%
\right] ^{1/2}  \label{modify10}
\end{eqnarray}%
while from Eq. (\ref{p21-a})%
\begin{eqnarray}
\left\vert C\right\vert &=&\frac{1}{M}\left[ \cos ^{2}\frac{\Delta _{\gamma }%
}{2}+\left( \frac{\Delta M}{M}\right) ^{2}\sin ^{2}\frac{\Delta _{\gamma }}{2%
}\right] ^{1/2}  \label{modify11} \\
&\simeq &\frac{1}{M}\left[ 1-\sin ^{2}\frac{\Delta _{\gamma }}{2}\right]
\notag
\end{eqnarray}%
neglecting $\left( \frac{\Delta M}{M}\right) ^{2}$ compared to $1$, which on
using Eq. (\ref{modify9}) becomes%
\begin{equation}
\left\vert C\right\vert \simeq \frac{1}{M}\left[ 1-\sin ^{2}2\theta
_{12}\sin ^{2}\left( \frac{\Delta }{2}\right) +O\left( \left( \frac{\Delta M%
}{M}\right) ^{2}\right) +O\left( \left( \frac{\Delta m}{m}\right)
^{2}\right) +O\left( s_{2}^{2}\right) \right] ^{1/2}  \label{modify12}
\end{equation}%
Thus%
\begin{eqnarray}
\left\vert h_{11}v_{2}\right\vert ^{2} &=&Mm\left[ 1-2\cos 2\theta _{12}%
\frac{\Delta m}{m}-\sin ^{2}2\theta _{12}\sin ^{2}\frac{\Delta }{2}\right]
^{1/2}\left[ 1-\sin ^{2}2\theta _{12}\sin ^{2}\frac{\Delta }{2}\right]
^{-1/2}  \notag \\
\left\vert 2h_{+}v_{1}\right\vert ^{2} &=&Mm\left[ 1+2\cos 2\theta _{12}%
\frac{\Delta m}{m}-\sin ^{2}2\theta _{12}\sin ^{2}\frac{\Delta }{2}\right]
^{1/2}\left[ 1-\sin ^{2}2\theta _{12}\sin ^{2}\frac{\Delta }{2}\right]
^{-1/2}  \notag \\
\left\vert 2h_{-}v_{1}\right\vert ^{2} &=&Mms_{2}^{2}\sin ^{2}2\theta _{12}
\left[ \sin ^{2}\frac{\Delta }{2}+\left( \frac{\Delta m}{m}\right) ^{2}%
\right] \left( 1-\sin ^{2}2\theta _{12}\sin ^{2}\frac{\Delta }{2}\right)
^{-1}\left[ 1+O\left( \frac{\Delta m}{m}\right) \right]  \label{modify13}
\end{eqnarray}

\section{Leptogenesis}

As is well known \cite{3, 4, 10, 11} the leptogenesis asymmetry is given by
\cite{11}
\begin{equation}
\epsilon _{i}=\frac{1}{8\pi }\sum_{k\neq i}\frac{1}{v_{i}^{2}R_{ii}}{Im}%
[(R_{ik})^{2}f(\frac{M_{k}^{2}}{M_{i}^{2}})]  \label{p35}
\end{equation}%
where $M_{i}$ denotes the heavy Majorana neutrino masses, $R_{ij}$ are
defined by%
\begin{equation}
R=\widehat{m}_{D}^{\dagger }\widehat{m}_{D}=V^{T}m_{D}^{\dagger
}m_{D}V^{\ast }  \label{p36}
\end{equation}%
The loop function $f(x)$ containing vertex and self-energy corrections is%
\begin{equation*}
\ f(x)=\sqrt{x}(\frac{2-x}{1-x}-(1+x)\ln \frac{1+x}{x})
\end{equation*}%
Now $\left( \left\vert v_{1}\right\vert ^{2}+\left\vert v_{2}\right\vert
^{2}\right) =(174GeV)^{2}=\left\vert v\right\vert ^{2}.$ One may take $%
\left\vert v_{1}\right\vert ^{2}=\left\vert v_{2}\right\vert ^{2}=\frac{1}{2}%
v^{2},$ so that
\begin{equation}
\epsilon _{1}=\frac{1}{8\pi }f\left( \frac{M_{2}^{2}}{M_{1}^{2}}\right)
\frac{1}{v_{1}^{2}R_{11}}{Im}[(R_{12})^{2}]  \label{p37}
\end{equation}%
Using the constraint \cite{2, 3}
\begin{equation}
R_{11}<4.3\times 10^{-7}v_{1}^{2},  \label{p38}
\end{equation}%
obtained from out of equilibrium decay of $M_{1}\simeq 10^{10}GeV,$one
finally obtains the lower limit on $\epsilon _{1}$:%
\begin{equation}
\epsilon _{1}=\frac{1}{8\pi }f\left( x\right) \frac{2.3\times 10^{6}}{%
v_{1}^{4}}\left\{ \Im\left[ \left( R_{12}\right) ^{2}\right] \right\}
\label{p39}
\end{equation}%
where, $x=\frac{M_{2}^{2}}{M_{1}^{2}}$, and for $M_{2}\simeq M_{1}$,
\begin{equation}
f(x)=-\frac{M}{4\Delta M}  \label{p40}
\end{equation}%
Now $\Im\left[ \left( R_{12}\right) ^{2}\right] $ as calculated from Eq. (%
\ref{p36}) is given by \cite{1}%
\begin{eqnarray}
\Im\left[ \left( R_{12}\right) ^{2}\right] &=&c^{\prime 2}s^{\prime
2}(\left\vert h_{11}v_{2}\right\vert ^{2}-\frac{1}{2}\left\vert
2h_{12}v_{1}\right\vert ^{2}+\left\vert 2h_{32}v_{1}\right\vert
^{2})^{2}\sin \Delta _{\gamma }  \notag \\
&=&\frac{1}{4}\left\{ \left\vert h_{11}v_{2}\right\vert ^{2}-\left\vert
2h_{+}v_{1}\right\vert ^{2}-\left\vert 2h_{-}v_{1}\right\vert ^{2}\right\}
^{2}\sin \Delta _{\gamma }  \label{modify16} \\
&=&M^{2}m^{2}\cos ^{2}2\theta _{12}\left( \frac{\Delta m}{m}\right)
^{2}\left\{ 1+\frac{1}{2}\frac{\sin ^{2}\theta _{13}}{\left( \Delta
m/m\right) }\frac{\sin ^{2}2\theta _{12}}{\cos 2\theta _{12}}\left( \sin ^{2}%
\frac{\Delta }{2}+\left( \frac{\Delta m}{m}\right) ^{2}\right) \right\} ^{2}%
\frac{1}{\left[ 1-\sin ^{2}2\theta _{12}\sin ^{2}\Delta /2\right] ^{2}}\sin
\Delta _{\gamma }  \notag
\end{eqnarray}%
Using Eqs. (\ref{modify8}), (\ref{p39}), (\ref{p40}) and (\ref{modify16})
along with%
\begin{equation}
\cos 2\theta _{12}=\frac{1}{3}\text{, }\sin ^{2}2\theta _{12}=\frac{8}{9}
\label{p41a}
\end{equation}%
\begin{eqnarray}
\epsilon &\simeq &\left( 6\times 10^{2}\right) \frac{M^{2}}{v_{1}^{4}}\Delta
m_{solar}^{2}\sin \Delta \left( \frac{\frac{\Delta m}{m}}{\frac{\Delta M}{M}}
\right) ^{2}\frac{1}{\left[ 1-\frac{8}{9}\sin ^{2}\frac{\Delta }{2}\right]
^{2}}\left\{ 1+\frac{4}{3}r^{\prime }\left( \sin ^{2}\frac{\Delta }{2}%
+\left( \frac{\Delta m}{m}\right) ^{2}\right) \right\} ^{2}\left[ 1+\frac{3}{
8}\left( 1-\frac{8}{9}\sin ^{2}\frac{\Delta }{2}\right) r^{\prime }\right]
\label{p42} \\
&\simeq &2\times 10^{-8}(\frac{M}{10^{10}GeV})^{2}\frac{\Delta m_{solar}^{2}
}{7.6\times 10^{-5}eV^{2}}(\frac{174GeV}{v})^{4}\sin \Delta \left( \frac{
\frac{\Delta m}{m}}{\frac{\Delta M}{M}}\right) ^{2}\times  \notag \\
&&\frac{1}{\left[ 1-\frac{8}{9}\sin^{2} \frac{\Delta }{2}\right] ^{2}}\left\{ 1+
\frac{4}{3}r^{\prime }\left( \sin ^{2}\frac{\Delta }{2}+\left( \frac{\Delta m
}{m}\right) ^{2}\right) \right\} ^{2}\left[ 1+\frac{3}{8}\left( 1-\frac{8}{9}
\sin ^{2}\frac{\Delta }{2}\right) r^{\prime }\right]  \label{p43}
\end{eqnarray}%
with $r^{\prime }=\sin ^{2}\theta _{13}/\left( \Delta m/m\right) $ , $%
v_{1}^{2}=\frac{1}{2}v^{2}=\frac{1}{2}(174GeV)^{2}$, $4m\Delta m=\Delta
m_{solar}^{2}$ and $\Delta m_{solar}^{2}=7.6\times 10^{-5}eV^{2}$. Using the
neutrino oscillation data \cite{12}
\begin{equation}
m\simeq (\Delta m_{atm}^{2})^{\frac{1}{2}}=4.9\times 10^{-2}eV
\label{p43constraint1}
\end{equation}%
\begin{eqnarray}
\frac{\Delta m}{m} &=&\frac{1}{4}\frac{\Delta m_{solar}^{2}}{\Delta
m_{atm}^{2}}=0.8\times 10^{-2}  \label{p43constraint2} \\
\sin ^{2}\theta _{13} &\leq &4.6\times 10^{-2}\left( 0.016\pm 0.010\right)
\label{p43constraint3}
\end{eqnarray}%
giving%
\begin{equation*}
r^{\prime }\leq 5.75\left( 2\pm 1.25\right)
\end{equation*}%
It can be seen from Eq. (\ref{p43}) that, the contribution from $\sin
^{2}\theta _{13}$ may dominate. The Majorana phase $\Delta $ which is the
same as would appear in neutrinoless double $\beta -$decay [c.f. first of
Eq. (\ref{modify7})] can be fixed from Eqs. (\ref{modify8}) and (\ref%
{modify9}). \ This gives%
\begin{equation*}
x\left[ \left( 1-x\right) +C\right] =r^{2}\left[ \left( 1-x\right) \left( x-%
\frac{1}{9}\right) \left( 1+\frac{3}{8}r^{\prime }x\right) ^{2}\right]
\end{equation*}%
where%
\begin{eqnarray*}
x &=&1-\frac{8}{9}\sin ^{2}\frac{\Delta }{2} \\
C &=&\left( \frac{\Delta M}{M}\right) ^{2}\left[ \frac{8}{9}r^{2}A-1\right]
\\
A &=&1+3r^{\prime }B\ ,\text{ \ \ \ \thinspace }1\leq A\leq 55 \\
B &=&1+\frac{3}{8}r^{\prime },~\text{\ ~\ \ \ }1\leq B\leq 3
\end{eqnarray*}%
for $r^{\prime }\leq 5.75$. It is clear that $\frac{1}{9}\leq x<1$. $C$ is
negligibly small except for $x\rightarrow 1$. In that case we have:

\textbf{Solution-I}

\begin{eqnarray}
\sin ^{2}\frac{\Delta }{2} &=&\frac{9}{8}\left( \frac{\Delta M}{M}\right)
^{2}\frac{\frac{8}{9}r^{2}A-1}{\frac{8}{9}r^{2}B^{2}-1}  \notag \\
\sin \Delta &=&\frac{3}{\sqrt{2}}\left( \frac{\frac{8}{9}r^{2}A-1}{\frac{8}{9%
}r^{2}B^{2}-1}\right) ^{1/2}\frac{\Delta M}{M}  \label{p43new}
\end{eqnarray}%
In this case, leptogenesis asymmetry $\epsilon $ given in Eq. (\ref{p42})
gives%
\begin{equation}
\epsilon \simeq 3\times 10^{-10}r\left( \frac{\frac{8}{9}r^{2}A-1}{\frac{8}{9%
}r^{2}B^{2}-1}\right) ^{1/2}B  \label{p43newzero}
\end{equation}%
Thus, $\epsilon $ is of right order of magnitude $\left(
10^{-6}-10^{-5}\right) $ for $1\leq \sqrt{A}\leq 7.4$ provided that $r\simeq
4\times 10^{3}$ i.e. $\frac{\Delta M}{M}\simeq 2\times 10^{-6}$.

\textbf{Solution-II}

There is another solution, for which for $r^{\prime }=0$, $\sin ^{2}\frac{%
\Delta }{2}=1-\frac{1}{8}\frac{1}{r^{2}-1}$, i.e. near maximal value as $%
\left( r^{2}-1\right) >1$. For $r^{\prime }\neq 0$, such a solution is
modified to

\begin{eqnarray}
\sin ^{2}\frac{\Delta }{2} &=&1-\frac{1}{8}\frac{1}{D^{2}r^{2}-1}  \notag \\
\sin \Delta &=&\frac{1}{\sqrt{2}}\frac{1}{rD}\left( 1-\frac{1}{r^{2}D^{2}}%
\right) ^{-1/2}\left[ 1-\frac{1}{8D^{2}r^{2}\left( 1-\frac{1}{D^{2}r^{2}}%
\right) }\right] ^{1/2}  \label{p43new1}
\end{eqnarray}%
where $D=1+r^{\prime }/24$. In this case%
\begin{equation*}
\epsilon \simeq 2\times 10^{-8}\frac{81}{\sqrt{2}}rf\left( r\right) \left[ 1+%
\frac{4}{3}\left( 1-\frac{1}{8}\frac{1}{D^{2}r^{2}-1}\right) r^{\prime }%
\right] ^{2}\left[ 1+\frac{1}{24D}\frac{1}{D^{2}r^{2}-1}r^{\prime }\right]
\end{equation*}%
where%
\begin{equation}
f\left( r\right) =\left( 1-\frac{1}{8\left( D^{2}r^{2}-1\right) }\right)
^{1/2}\left( 1-\frac{1}{D^{2}r^{2}}\right) ^{3/2}  \label{p47}
\end{equation}%
The asymmetry $\epsilon $ and $\sin ^{2}\Delta /2$ are plotted as a function
of $r$ for $r^{\prime }=0,$ $2$ and $5.75$ in Figure 1(a,b,c) and Fig. 2
respectively.
\begin{figure}[tbp]
\begin{center}
\begin{tabular}{ccc}
\vspace{-2cm} \includegraphics[scale=0.99]{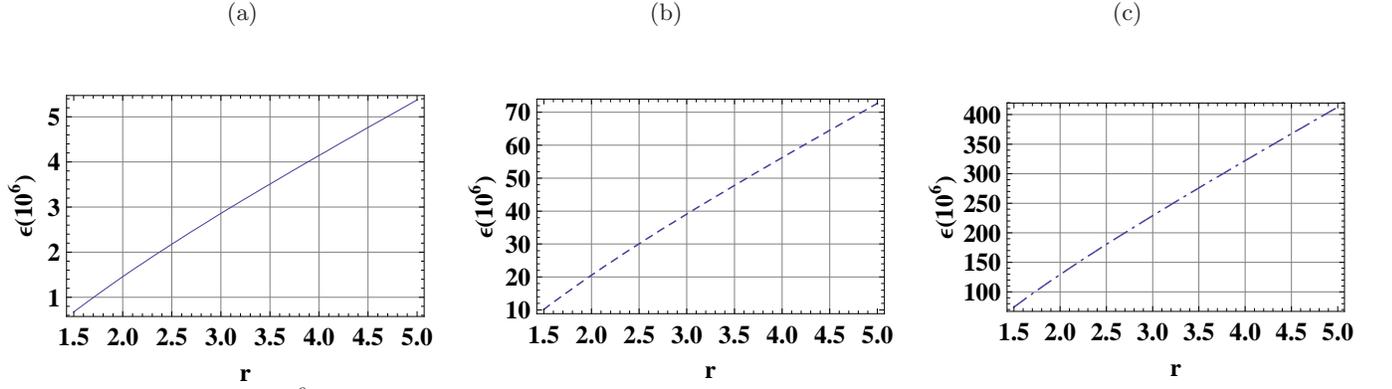} \put (-435,170){(a)}
\put (-275,170){(b)}\put (-100,170){(c)} &  &
\end{tabular}%
\end{center}
\caption{The asymmetry $\protect\epsilon \times 10^{6}$ as a function of $r$
for different values of $r^{\prime}$. The labels a, b and c corresponds to $%
r^{\prime}=0$, $2$ and $5.75$, respectively.}
\end{figure}

\begin{figure}[tbp]
\begin{center}
\vspace{-2cm} \includegraphics[scale=0.99]{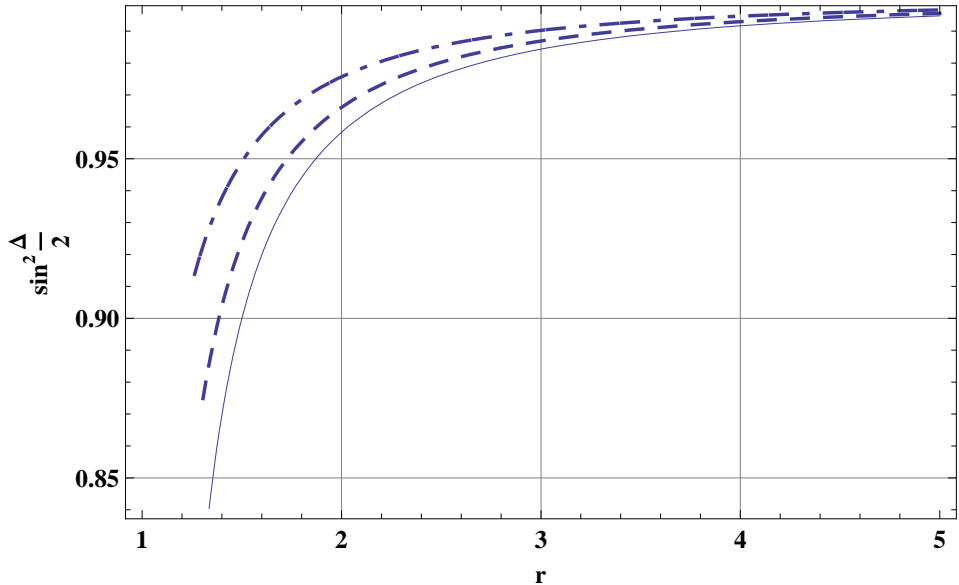}
\end{center}
\caption{$\sin ^{2}\Delta /2$ as a function of $r$ for different values of $%
r^{\prime}$. Solid line correspond to $r^{\prime}$=0, dashed line is for $%
r^{\prime}$=2, and double-dashed-dotted is for $r^{\prime}$=5.75.}
\end{figure}
One can see from the plot that (i) $\sin ^{2}\Delta /2$ is near maximal ($%
\geq 0.90$) for any \thinspace $r\geq 1.5$ and $r^{\prime }$. (ii) For any
given value of $r$, $\sin ^{2}\theta _{13}$ gives the dominant contribution
to $\epsilon $, e.g. for $r=2$, $\epsilon $ is $1.5\times 10^{-6}$ $\left(
r^{\prime }=0\right) $, $2.0\times 10^{-5}\left( r^{\prime }=2\right) $ and $%
1.3\times 10^{-4}\left( r^{\prime }=5.75\right) $. (iii) An accurate
measurement of $r^{\prime }$ will have important implications for $r$ and in
turn for $\frac{\Delta M}{M}$. The value $r\simeq 2$ implies $\frac{\Delta M%
}{M}\simeq 4\times 10^{-3}$ which gives the degeneracy required for heavy
right handed neutrinos. It is important to note that CP violation
responsible for the generation of baryogenesis parameter through
leptogenesis comes entirely from Majorana phase $\Delta $ which is now
predicted to be negligible for the first solution and for the second
solution $\sin ^{2}\frac{\Delta }{2}\succeq 0.95$. This can in principle be
tested in neutrinoless double $\beta -$decay, where the effective electron
neutrinos mass is given by%
\begin{eqnarray}
m_{ee} &=&\left\vert a\right\vert  \notag \\
&\simeq &m\left[ 1-\sin ^{2}2\theta _{1}\sin ^{2}\frac{\Delta }{2}\right]
^{1/2}  \notag \\
&\simeq &4.3\times 10^{-2}\text{eV, for solution I}  \label{p48} \\
&\simeq &1.7\times 10^{-2}\text{eV, for solution II}  \label{p49}
\end{eqnarray}

\section{Conclusion}

By considering a simple generic seesaw model gauge model with $\mu -\tau $
symmetry for the heavy right-handed neutrinos, degenerate leptogenesis has
been studied, where the exact degeneracy is due to $\bar{L}$ discrete
symmetry for the heavy right handed neutrinos. When this degeneracy is
slightly broken, an adequate lepton asymmetry $\left( \epsilon \simeq
10^{-6}-10^{-5}\right) $ can be obtained. The level of degeneracy required
in one case is $\frac{\Delta M}{M}\simeq 10^{-6}$, much smaller than $\frac{%
\Delta m}{m}\simeq 8\times 10^{-3}$ obtained from neutrino oscillations and
in the second case is $\frac{\Delta M}{M}\simeq 4\times 10^{-3}$ which is of
the same order as $\frac{\Delta m}{m}\simeq 8\times 10^{-3}$. This in turn
predicts the Majorana phase responsible for the lepton asymmetry. Since the
same phase appear in neutrinoless double $\beta -$decay, it can in principle
be tested. Further $\sin ^{2}\theta _{13}$ has important implications for
the leptogenesis asymmetry parameter $\epsilon $ and degree of degeneracy $%
\frac{\Delta M}{M}$, needed. Thus an accurate measurement of $\sin
^{2}\theta _{13}$ will be of great interest.\\
\textbf{Note:} I have been informed by Werner Rodejohann about the reference \cite{13} where the various consequences of {$\left( L_{e}-L_{\mu }-L_{\tau }\right) $ discrete symmetry for the light neutrino sector, including degenerate leptogenesis, has been discussed in a different context.\\
\textbf{Acknowledgement:} The author would like to thank Prof. K.
Sreenivasan for hospitality at Abdus Salam International Centre for
Theoretical Physics, Trieste where a part of this work was done.

\end{document}